%% file: main4.tex
\documentclass[letterpaper,11pt]{article}
\pdfoutput=1
\usepackage{braket}
\usepackage{jheppub}

\usepackage{tikz}
\usepackage{array}
\usetikzlibrary{positioning}
\usetikzlibrary{shapes.geometric}
\usetikzlibrary{arrows}

\newcommand{\an}[1]{\left\langle#1\right\rangle}
\newcommand{\e}{\epsilon}

\newcolumntype{L}{>{$}l<{$}}

\title{Landau Singularities of the 7-Point Ziggurat II}

\author[a]{Luke Lippstreu,}
\author[a,b]{Marcus Spradlin,}
\author[a]{Akshay Yelleshpur Srikant}
\author[a]{and Anastasia Volovich}

\affiliation[a]{Department of Physics,
	Brown University,
	Providence, RI 02912, USA}
\affiliation[b]{Brown Theoretical Physics Center,
	Brown University,
	Providence, RI 02912, USA}

\emailAdd{luke\_lippstreu@brown.edu}
\emailAdd{marcus\_spradlin@brown.edu}
\emailAdd{akshay\_yelleshpur\_srikant@brown.edu}
\emailAdd{anastasia\_volovich@brown.edu}

\abstract{We solve the Landau equations to find the singularities of nine three-loop 7-point graphs that arise as relaxations of the graph studied in~arXiv:2211.16425. Along the way we establish that $Y{-}\Delta$ equivalence fails for certain branches of solutions to the Landau equations. We find two graphs with singularities outside the heptagon symbol alphabet; in particular they are not cluster variables of ${\rm Gr}(4,7)$. We compare maximal residues of scalar graphs exhibiting these singularities to those in $\mathcal{N}=4$ super-Yang-Mills theory in order to probe their cancellation from its amplitudes.}

\begin{document}

\maketitle

\section{Introduction}

Scattering amplitudes are heavily constrained by their analytic structure and satisfy many remarkable properties in the vicinity of their singular points~\cite{Landau:1959fi, ELOP}. Cataloging these singular points remains an important open problem which has received a lot of recent attention, see for example~\cite{Bourjaily:2020wvq, Hannesdottir:2021kpd, Correia:2021etg, Hannesdottir:2022xki, Mizera:2022dko, Berghoff:2022mqu, Mizera:2021icv, Bourjaily:2022vti, Flieger:2022xyq, He:2022tph, Chen:2023kgw, Dlapa:2023cvx, Prlina:2018ukf} and also~\cite{Prlina:2017tvx, Prlina:2017azl, He:2020uhb, Dennen:2015bet, Dennen:2016mdk, Mago:2020kmp} for work specific to planar $\mathcal{N}=4$ super Yang-Mills (SYM) theory.

This paper is a companion to~\cite{Lippstreu:2022bib}, to which we refer the reader for a more thorough introduction. The goal of this work is to investigate what can be said about the singularity structure of massless planar 7-point Feynman integrals in arbitrary (four-dimensional) quantum field theories, using the Landau equations~\cite{Landau:1959fi}. A primary motivation for this work is to provide a better foundation for the amplitude bootstrap program in $\mathcal{N}=4$ SYM theory (see~\cite{Caron-Huot:2020bkp, Papathanasiou:2022lan, Arkani-Hamed:2022rwr} for reviews), which represents the state of the art for multi-loop computation and starts from the assumption (supported by all evidence available to date) that its 7-point amplitudes only have singularities at the vanishing locus of certain kinematic invariants called the \emph{heptagon symbol alphabet}. This alphabet, which first appeared in~\cite{Caron-Huot:2011zgw}, consists of 49 symbol letters~\cite{Goncharov:2010jf} falling into seven cyclic families $b_0, \ldots, b_6$ with representatives
\begin{align}
\label{eq:heptagonalphabet}
    &b_0 \ni  \an{1234}, \,\,  b_1 \ni \an{1256}, \,\, b_2 \ni \an{1456}, \,\, b_3 \ni \an{1236},\\
    &\nonumber b_4 \ni \an{1346}, \,\,  b_5 \ni \an{1(23)(45)(67)}, \,\, b_6 \ni \an{1(34)(56)(72)},
\end{align}
where $\an{a(bc)(de)(fg)} = \an{abde}\an{acfg}-\an{abfg}\an{acde}$ using momentum twistor notation. It was pointed out in~\cite{Golden:2013xva} that these 49 letters are precisely the cluster variables of the ${\rm G}(4,7)$ cluster algebra.

In~\cite{Lippstreu:2022bib} the authors determined the leading\footnote{Here ``leading Landau singularity'' means solutions of the Landau equations which have all propagators associated to a graph on-shell regardless of the vanishing of the corresponding Feynman parameter. Similarly, ``subleading Landau singularity'' refers to solutions in which all propagators but one are on-shell, etc. It should be noted that while this usage has been employed in previous works by some of the authors of this paper~\cite{Dennen:2015bet, Dennen:2016mdk, Prlina:2017azl, Prlina:2017tvx, Prlina:2018ukf, Lippstreu:2022bib}, it differs slightly from usage in older literature on this subject.\label{footnote1}} (first-type) Landau singularities of a certain special four-loop 7-point graph $\mathcal{G}_7$ (shown in Fig.~2(b) of that paper) with the property that any \emph{other} planar 7-point graph, at any loop order, can be reduced to $\mathcal{G}_7$ (or one of its relaxations) by some sequence of graphical moves familiar from electrical circuit theory, in particular including the $Y{-}\Delta$ (also called star-triangle) move. (The graph $\mathcal{G}_7$ itself is related by such moves to a graph called the ``ziggurat'' in~\cite{Prlina:2018ukf}.) This was motivated by the argument in~\cite{Dennen:2016mdk,Prlina:2018ukf} that the locus of solutions of the Landau equations is invariant under this set of moves. It was found in~\cite{Lippstreu:2022bib} that $\mathcal{G}_7$ has singularities at the vanishing of letters in all cyclic classes except $b_6$, and no singularities outside of the heptagon alphabet. It was pointed out that relaxations of $\mathcal{G}_7$ certainly have singularities of type $b_6$, and left open for the present work the question of whether relaxations of $\mathcal{G}_7$ might have any additional singularities \emph{outside} the heptagon alphabet.

In this paper we extend the analysis of~\cite{Lippstreu:2022bib} by solving the Landau equations for several three-loop\footnote{In principle one must also look at two- and one-loop relaxations, but those graphs are all sufficiently simple that it is quick to see that they do not yield any singularities outside the heptagon alphabet.} relaxations of $\mathcal{G}_7$ (shown in Figures~\ref{fig:graph1}--\ref{fig:graph2} and~\ref{fig:firstfig}--\ref{fig:lastfig}). Along the way, we establish that there are branches of solutions that evade the analysis of~\cite{Prlina:2018ukf}, and conclude that the $Y{-}\Delta$ equivalence of the Landau equations is false. This failure can occur for graphs having a 3-point vertex bounded by three internal faces, so the simplest examples start at three-loop order with graphs of the ``tennis court'' topology. Figures~\ref{fig:graph1} and~\ref{fig:graph2} show two examples of 7-point graphs for which we establish the presence of singularities (see equations~(\ref{eq:classc0}) and (\ref{eq:classc1}), respectively) outside of the heptagon 
 alphabet. (Note however that Figures~\ref{fig:firstfig}, \ref{fig:graph5} and~\ref{fig:graph6} show graphs with an internal vertex which do not have such singularities.)

The structure of this paper is as follows.  In Section~\ref{sec:YDelta} we explain the subtle branch of solutions to the Landau equations which can cause $Y{-}\Delta$ equivalence to fail for graphs having an entirely internal 3-vertex.  In Section~\ref{sec:newletters} we demonstrate the presence of singularities outside the heptagon alphabet for two particular three-loop 7-point graphs.  Since it is known that these singularities are absent from the three-loop 7-point MHV~\cite{Drummond:2014ffa} and NMHV~\cite{Dixon:2016nkn} amplitudes in SYM theory, we investigate in Section~\ref{sec:five} the cancellation of these singularities by comparing some maximal residues\footnote{These are usually called ``leading singularities'' in the literature, and we follow this convention in Section~\ref{sec:five}, but we acknowledge here that this can be a source of some confusion.} of scalar integrals compared to those in $\mathcal{N}=4$ SYM theory. We end in Section~\ref{sec:discussion} with a discussion of the implications of our results and some open questions.

\section{\texorpdfstring{$Y{-}\Delta$ inequivalent configurations}{Y-Delta inequivalent configurations}}
\label{sec:YDelta}

In this section, we will demonstrate that there are certain configurations of graphs for which $Y{-}\Delta$ equivalence of (the solution locus of) the Landau equations fails.
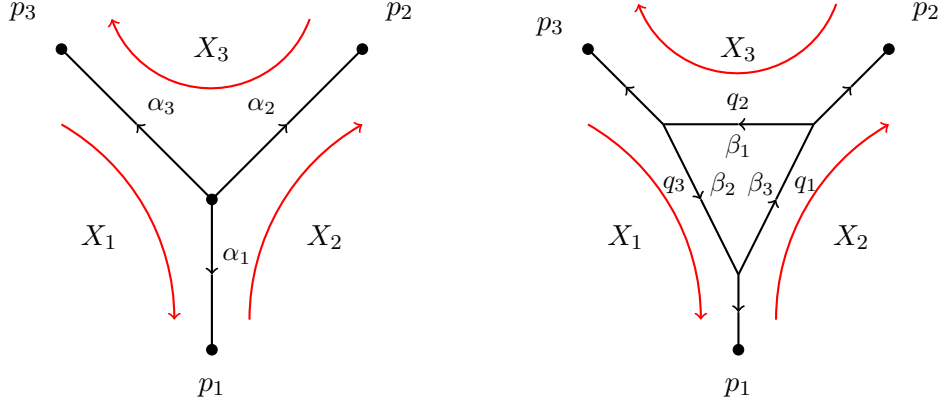
\begin{figure}[htb!]
    \centering
    \input{YDelta.tex}
\end{figure}

Let us begin by analyzing the solutions to the $Y$ system shown on the left, which is generically embedded inside an arbitrary larger planar graph. The outgoing momenta must of course satisfy momentum conservation
\begin{align}
    \label{eq:momconservation}
    p_1 + p_2 + p_3 = 0\,.
\end{align}
The Landau equations for this system are the on-shell conditions\footnote{Note that we always impose the on-shell condition on each internal momentum regardless of whether the corresponding $\alpha_i$ is vanishing; see footnote~\ref{footnote1} and the discussion in~\cite{Prlina:2018ukf}.}
\begin{align}
    \label{eq:Y-Landau - O}
    p_1^2 =  p_2^2 =p_3^2 =0
\end{align}
and the Kirchhoff equations\footnote{We suppress Lorentz four-vector indices on the $p_i$ and $X_i$.}
\begin{align}
 &\nonumber-\alpha_1 p_1 + \alpha_3 p_3 + X_1 = 0\,,\\
    &\label{eq:Y-Landau - K}  -\alpha_1 p_1 + \alpha_2 p_2 - X_2 = 0\,,\\
    &\nonumber -\alpha_2 p_2 + \alpha_3 p_3 - X_3 = 0\,,
\end{align}
where $X_1, X_2, X_3$ represent contributions from the rest of the graph. Note that this system is consistent only if
\begin{align}
    \label{eq:Y-system consistency}
    X_1 + X_2 + X_3 = 0\,.
\end{align}
If we assume that the $\alpha_i$ are all non-zero\footnote{The Landau equations for massless graphs often have interesting and relevant solutions when one or more of the $\alpha$'s are zero, but our purpose in this section is merely to provide \emph{an} example where $Y{-}\Delta$ equivalence fails, and for that purpose, it suffices to look at an example where all three $\alpha_i$ are nonzero.}, the solution to~(\ref{eq:Y-Landau - K}) is
\begin{align}
    \label{eq:Ysol-K}
    p_1 = \frac{\alpha_3}{\alpha_1}p_3 + \frac{1}{\alpha_1} X_1\,, \qquad p_2 = \frac{\alpha_3}{\alpha_2}p_3 - \frac{1}{\alpha_2} X_3\,.
\end{align}
Furthermore, momentum conservation~(\ref{eq:momconservation}) requires
\begin{align}
\label{eq:pthree}
    \left(\alpha_1 \alpha_2 + \alpha_2 \alpha_3 + \alpha_1 \alpha_3\right) p_3 + \alpha_2 X_1 - \alpha_1 X_3 = 0\,.
\end{align}
Generically, equation~(\ref{eq:pthree}) (and its three cyclic images) determine
\begin{align}
    \label{eq:Ysol-genericX}
    p_1 &= \frac{\alpha_2 X_1 - \alpha_3 X_2}{\alpha_1 \alpha_2 + \alpha_2 \alpha_3 + \alpha_1 \alpha_3 }\,,\nonumber\\
    p_2 &= \frac{\alpha_3 X_2 - \alpha_1 X_3}{\alpha_1 \alpha_2 + \alpha_2 \alpha_3 + \alpha_1 \alpha_3 }\,,\\
    p_3 &= \frac{\alpha_1 X_3 - \alpha_2 X_1}{\alpha_1 \alpha_2 + \alpha_2 \alpha_3 + \alpha_1 \alpha_3}\,.\nonumber
\end{align}
However, in the special case $X_1 = X_2 = X_3 = 0$ there is another branch of solutions
\begin{align}
 \label{eq:Ysol-zeroX}
    \alpha_1 \alpha_2 &+ \alpha_2 \alpha_3 + \alpha_1 \alpha_3  = 0\,, \qquad p_1 = \frac{\alpha_3}{\alpha_1} p_3\,, \qquad p_2 = \frac{\alpha_3}{\alpha_2} p_3\,,
\end{align}
with $p_3$ being on-shell ($p_3^2 = 0$) but otherwise arbitrary. Since $p_1, p_2, p_3$ are all collinear, we also have $p_1^2 = p_2^2 = 0$.

We will now demonstrate that the solutions of the corresponding $\Delta$ system (shown above on the right) are equivalent only away from the branch~(\ref{eq:Ysol-zeroX}) where all $X_i$ are zero. The Landau equations for the $\Delta$ system are the on-shell conditions (using momentum conservation at every vertex gives $q_2 =q_1-p_2, q_3 = p_1 + q_1$)
\begin{align}
    \label{eq:Delta-Landau-O}
    (q_1 - p_2)^2 = (p_1+q_1 )^2 =  q_1^2 = 0
\end{align}
and the Kirchhoff equations
\begin{align}
    \label{eq:Delta-Landau-K1}  &q_1 \sum_{i}\beta_i -\beta_1 p_2 + \beta_2 p_1 = 0\,,\\
    \label{eq:Delta-Landau-K2}  &\beta_1 (q_1 - p_2) - X_3=  \beta_2(p_1+q_1 ) -X_1= \beta_3 q_1 - X_2 = 0\,.
\end{align}
Solving (\ref{eq:Delta-Landau-K1}) for $q_1$ gives
\begin{align}
\label{eq:Delta-qsol}
    q_1 = \frac{\beta_1 p_2 - \beta_2 p_1}{\beta_1+\beta_2+\beta_3}\,.
\end{align}
When the $X_i$ are not all zero, we can set\footnote{This is the usual relation between the resistances of an electrical circuit network under $Y{-}\Delta$ equivalence.}
\begin{align}
    \label{eq:Delta-Y map}
    \beta_i = \frac{\alpha_1 \alpha_2 + \alpha_2 \alpha_3 + \alpha_1 \alpha_3}{\alpha_i}
\end{align}
to map solutions of~(\ref{eq:Y-Landau - K}) to solutions of~(\ref{eq:Delta-Landau-K2}), effectively solving (\ref{eq:Delta-Landau-K1}--\ref{eq:Delta-qsol}). We are left with the on-shell conditions~(\ref{eq:Delta-Landau-O}) which admit a non-trivial solution (this means the three $\beta_i$'s are not all zero) only when
\begin{align}
    p_1^2 = p_2^2 = p_3^2 = 0\,.
\end{align}
Thus we recover the on-shell conditions of~(\ref{eq:Delta-Landau-O}) and thereby establish the equivalence of the two systems.

However, for the special case $X_1 = X_2 = X_3 = 0$ the above map breaks down since we see from~(\ref{eq:Ysol-zeroX}) and~(\ref{eq:Delta-Y map}) that it would set all three $\beta_i$ to zero. Moreover, we can certify that the solution~(\ref{eq:Ysol-zeroX}) of the $Y$ system does not correspond to any solution in the $\Delta$ system by tabulating all solutions of the latter:
\begin{enumerate}
    \item $\beta_1 = 0, \beta_2 = 0, \beta_3 = 0$, \qquad  $p_1^2 \neq 0, p_2^2 \neq 0, p_3^2 \neq 0$,
    \item $\beta_1 = 0, \beta_2 \neq 0, \beta_3 \neq 0$, \qquad  $p_1 = 0 , p_2^2 = 0, p_3^2 = 0$,
    \item $\beta_1 \neq 0, \beta_2 = 0, \beta_3 \neq 0$, \qquad  $p_1^2 = 0 , p_2 = 0, p_3^2 = 0$,
    \item $\beta_1 \neq 0, \beta_2 \neq 0, \beta_3 = 0$, \qquad  $p_1^2 = 0 , p_2^2 = 0, p_3^2 \neq 0$,
    \item $\beta_1 = 0, \beta_2 = 0, \beta_3 \neq 0$, \qquad  $p_1^2 = 0 , p_2^2 = 0, p_3^2 \neq 0$,
    \item $\beta_1 = 0, \beta_2 \neq0, \beta_3 = 0$, \qquad  $p_1^2 = 0 , p_2^2 \neq 0, p_3^2 = 0$,
    \item $\beta_1 \neq 0, \beta_2 = 0, \beta_3 = 0$, \qquad  $p_1^2 \neq 0 , p_2^2 = 0, p_3^2 = 0$,
    \item $\beta_1 \neq 0, \beta_2 \neq 0, \beta_3 \neq 0$, \qquad  $p_1 = 0 , p_2 = 0, p_3 = 0$.
\end{enumerate}

None of these solutions have all three momenta collinear, on-shell and non-zero, like the solution~(\ref{eq:Ysol-zeroX}). We conclude that there are branches of solutions to the Landau equations for which the $Y{-}\Delta$ equivalence breaks down. These branches are somewhat subtle, requiring the presence of an internal 3-point vertex at which $X_1 = X_2 = X_3 = 0$. For this reason, we are not aware of any examples of this breakdown having been noted in the previous literature; we encountered this phenomenon in solving the Landau equations for the various three-loop graphs analyzed in the next section.

\section{Letters beyond the heptagon alphabet}
\label{sec:newletters}

In this section, we will analyze the solutions of the Landau equations for two particular three-loop graphs that arise as relaxations of the four-loop graph studied in~\cite{Lippstreu:2022bib}. These two are of particular interest because they both admit solutions whose existence requires the vanishing of kinematic invariants outside the heptagon symbol alphabet. These solutions occur for configurations that are precisely of the $Y{-}\Delta$ violating type described in the previous section. Results for several other three-loop graphs that do not have singularities outside the heptagon alphabet are summarized in the Appendix.

While solving the Landau equations, we will exploit simplifying features of momentum twistor and loop momentum space as appropriate. For instance, solving the on-shell conditions is easiest in momentum twistor space where it amounts to tabulating solutions to various Schubert problems~\cite{Arkani-Hamed:2010pyv, Morales:2022csr, He:2022tph, Yang:2022gko}. In contrast, solving the Kirchhoff equations is more convenient using standard momentum variables. We can construct the momentum 4-vector corresponding to twistors $A, B, C$ and $D$ via the map~\cite{Mason:2009qx}
\begin{align}
\label{eq:tvmap}
    p_{a \dot{a}} \left(A, B, C, D\right) = \mathcal{I}^{\alpha \beta} \mathcal{I}_{\gamma \delta} \frac{\epsilon_{\beta}\left(\cdot, A, B, C\right)D^{\delta} - \epsilon_{\beta}\left(\cdot, A, B, D\right)C^{\delta}}{\an{AB}\an{CD}}
\end{align}
where $I^{\alpha\beta}, I_{\gamma\delta}$ refer to the infinity twistor and its dual and we use the shorthand $\epsilon(\cdot,A,B,C)=\e^{\mu\nu\rho\sigma}A_{\nu}B_{\rho}C_{\sigma}$, with Greek indices running from $0$ to $3$ and Latin indices from $1$ to $2$. We refer the reader to~\cite{Dennen:2016mdk, Prlina:2017azl, Prlina:2017tvx, Lippstreu:2022bib} for more details on solving Landau equations using momentum twistors.

\subsection{Graph 1}

The first graph of interest is displayed in Fig.~\ref{fig:graph1}. The 12 on-shell conditions
\begin{align}
    &\label{eq:g1-ll} \an{ABEF} = \an{ABCD} = \an{EFCD} = 0\,,\\
    &\label{eq:g1-ab} \an{AB17} = \an{AB67} = \an{AB56} = 0\,,\\
    &\label{eq:g1-cd} \an{CD12} = \an{CD23} = \an{CD34} = 0\,,\\
    &\label{eq:g1-ef} \an{EF34} = \an{EF45} = \an{EF56} = 0\
\end{align}
completely localize the loop momenta. The first three equations, in~(\ref{eq:g1-ll}), involve only loop momenta and have two solutions~\cite{Arkani-Hamed:2018rsk, Langer:2019iuo, Dian:2022tpf}. The first is a configuration of lines intersecting at a single arbitrary point $X$:
\begin{align}
    \label{eq:all-in-point}
    AB = (X,P_1)\, \qquad CD = (X,P_2)\,, \qquad EF=(X,P_3)\,,
\end{align}
with the $P_i$ being three arbitrary points distinct from $X$. The second is the parity conjugate of the above solution, given by a configuration of three coplanar but otherwise generic lines. We begin by focusing only on the class of solutions given by~(\ref{eq:all-in-point}). The remaining constraints~(\ref{eq:g1-ab}--\ref{eq:g1-ef}) then further restrict $X, P_1, P_2, P_3$ to satisfy
\begin{align}
    &\label{eq:g1-ab-sols} XP_1 =
    \begin{cases}
         \left(167\right) \cap \left(X56\right) \qquad \an{X167} = 0\,,\\
         \left(X17\right) \cap \left(567\right) \qquad \an{X567} = 0\,,
    \end{cases}\\
     &\label{eq:g1-cd-sols} XP_2 =
     \begin{cases}
         \left(123\right) \cap \left(X34\right) \qquad \an{X123} = 0\,,\\
         \left(X12\right) \cap \left(234\right) \qquad \an{X234} = 0\,,
    \end{cases}\\
     &\label{eq:g1-ef-sols} XP_3 =
     \begin{cases}
         \left(X34\right) \cap \left(456\right) \qquad \an{X456} = 0\,,\\
         \left(345\right) \cap \left(X56\right) \qquad \an{X345} = 0\,.
    \end{cases}
\end{align}
Altogether there are 8 solutions. For example, picking the first line of each of~(\ref{eq:g1-ab-sols}--\ref{eq:g1-ef-sols}) gives
\begin{equation}
\begin{aligned}
    \label{eq:g1-example-sol}
    X &= \left(167\right) \cap \left(123\right) \cap \left(456\right)\,,\\
    XP_1 &= \left(167\right) \cap \left(456\right)\,,\\
    XP_2 &= \left(123\right) \cap \left(X34\right)\,,\\
    XP_3 &= \left(X34\right) \cap \left(456\right)\,.
\end{aligned}
\end{equation}
The parity conjugate of this solution can be written compactly as
\begin{align}
\label{eq:g1-example-sol-pc}
    \overline{X} = (725)\,, \quad \overline{XP_1} = (7,5)\,,\quad \overline{XP_2} = (2, 725 \cap 35)\,, \quad \overline{XP_3} = (5, 725\cap 34)\,.
\end{align}
The remaining seven solutions can be worked out in a similar manner and we summarize the on-shell solutions in Fig.~\ref{fig:graph1}.

\begin{figure}
\centering
  \begin{minipage}[c]{0.45\textwidth}
      \centering
    \input{graph1.tex}
  \end{minipage}
  \hfill
  \begin{minipage}[c]{0.9\textwidth}
    \centering
     \begin{tabular}{|l|l|L | L | L|L|}
        \# & $\overline{X}$ & \overline{AB} & \overline{CD} & \overline{EF} & \text{Letter classes}\\ \hline \hline
         1 & (725) & \left(7,5\right) & (2,725\cap 35) & (5,725 \cap 34) & b_0, b_3, b_5\\
         2 & (724) & \left(7,724\cap 56\right) & (2,4) & (4,724 \cap 56) & b_0, b_3, b_5\\
         3 & (735) & \left(7,5\right) & (3,735\cap 12) & (5,3) & b_0, b_2, b_3, b_5\\
         4 & (734) & \left(7,734\cap 56\right) & (3,734\cap 12) & (4,734 \cap 56)  & b_0, b_1, b_2, b_3, \bar{c}_0\\
         5 & (625) & \left(6,625\cap 17\right) & (2,625\cap 34) & (5,625 \cap 34)& b_0, b_1, b_2, b_3, c_0\\
         6 & (624) & \left(6,624\cap 17\right) & (2,4) & (4,6) & b_0, b_2, b_3, b_5\\
         7 & (635) &\left(6,635\cap 17\right) &(3,635\cap 12) & (5,3) & b_0, b_1, b_2, b_3\\
         8 & (634) & \left(6,634\cap 17\right) &(3,634\cap 12) & (4,6) & b_0, b_1, b_2, b_3\\ \hline
         \end{tabular}
    \end{minipage}
    \caption{Cuts and first-type Landau singularities, up to parity conjugates, for the graph shown.}
    \label{fig:graph1}
\end{figure}

Next, we must solve the Kirchhoff conditions for each on-shell solution listed.  Because the graph has 12 internal edges, each with an associated Feynman parameter $\alpha_i$, the Kirchhoff conditions can be expressed as a $12 \times 12$ system of homogeneous equations in the $\alpha_i$. Non-trivial solutions exist only when the associated determinant, which is a product of kinematic invariants, vanishes. This determinant is the locus on which the integral can develop singularities and each factor in this determinant is a candidate for a symbol letter. While most of the solutions lead only to letters in the heptagon symbol alphabet~(\ref{eq:heptagonalphabet}), solutions \#4 and \#5 in the table (and their parity conjugates) produce new letters outside the heptagon alphabet. Interestingly all of these solutions arise from $Y{-}\Delta$ violating configurations of the type described in Section~\ref{sec:YDelta}.

For illustration, we present the full details for the solution corresponding to the parity conjugate of cut \#4 in Fig.~\ref{fig:graph1}, which is given by
\begin{gather}
AB=(6,X)\,, \quad CD=(2,X)\,, \quad EF=(5,X)\,, \quad X = (34)\cap\bar{7}\,.
\end{gather}
Using this on-shell solution and the map~(\ref{eq:tvmap}), the momenta $\ell_{\text{\tiny{ABCD}}}, \ell_{\text{\tiny{CDEF}}}$ and $\ell_{\text{\tiny{ABEF}}}$ of the lines corresponding to the Feynman parameters $\alpha_{10}, \alpha_{11}$ and $\alpha_{12}$ respectively are\footnote{Here we have once again suppressed all indices.}
\begin{gather}
    \ell_{\text{\tiny{ABCD}}}=\frac{\e(\cdot,6,X,2)}{\braket{AB}\braket{CD}}X\,,\quad \ell_{\text{\tiny{ABEF}}}=\frac{\e(\cdot,6,X,5)}{\braket{AB}\braket{EF}}X\,,\quad \ell_{\text{\tiny{CDEF}}}=\frac{\e(\cdot,2,X,5)}{\braket{CD}\braket{EF}}X\,.
\end{gather}
A particular solution to the Kirchhoff equations is
\begin{align}
    \alpha_1 = \cdots = \alpha_9 = 0
\end{align}
with the remaining Feynman parameters constrained by
\begin{gather}
    \nonumber \alpha_{10} \, \ell_{\text{\tiny{ABCD}}} + \alpha_{11}\,\ell_{\text{\tiny{CDEF}}}=0\,,\\
     \alpha_{10} \, \ell_{\text{\tiny{ABCD}}} + \alpha_{12} \,\ell_{\text{\tiny{ABEF}}} = 0\,,\\
    \nonumber \alpha_{12}\, \ell_{\text{\tiny{ABEF}}} - \alpha_{11} \,\ell_{\text{\tiny{CDEF}}} = 0\,.
\end{gather}
This system of equations has a solution only if
\begin{gather}
    \braket{256X} =  \braket{256(34)\cap(671)}=  \braket{6(17)(25)(34)} = 0\,.
\end{gather}
The letter doesn't belong to the heptagon alphabet. Considering the cyclic permutations of this graph, together with the parity conjugate of this cut, generates two new cyclic families of letters containing the representatives
\begin{gather}
    \label{eq:classc0}
    c_0 \ni  \braket{1(23)(47)(56)}\,, \qquad\bar{c}_0 \ni \an{1(25)(34)(67)}\,.
\end{gather}
The various cyclic classes of letters encountered in the Landau analysis of all cuts of this graph are summarized in the final column of the table in Fig.~\ref{fig:graph1}.

\subsection{Graph 2}

The second graph of interest is shown in Fig.~\ref{fig:graph2}. Many aspects of its Landau analysis differ significantly from graph 1. These differences stem from the fact that this graph only has 11 propagators. Consequently, the on-shell conditions
\begin{align}
    &\label{eq:g2-ll} \an{ABEF} = \an{ABCD} = \an{EFCD} = 0\,,\\
    &\label{eq:g2-ab} \an{AB17} = \an{AB12} = 0\,,\\
    &\label{eq:g2-cd} \an{CD23} = \an{CD34} = \an{CD45} = 0\,,\\
    &\label{eq:g2-ef} \an{EF45} = \an{EF56} = \an{EF67} = 0
\end{align}
do not localize the loop momenta completely and all cut solutions have one residual degree of freedom. As before, the solutions to~(\ref{eq:g2-ll}) are
\begin{align}
    AB = (X,P_1)\,, \qquad CD = (X,P_2)\,, \qquad EF=(X,P_3)
\end{align}
and its parity conjugate. This solves~(\ref{eq:g2-ab}--\ref{eq:g2-ef}) if
\begin{align}
    &\label{eq:g2-ab-sols} XP_1 = \left(X12\right) \cap \left(X17\right) = (X,1)\,,\\
     &\label{eq:g2-cd-sols} XP_2 =
     \begin{cases}
         \left(234\right) \cap \left(X45\right) \qquad \an{X234} = 0\,,\\
         \left(X23\right) \cap \left(345\right) \qquad \an{X345} = 0\,,
    \end{cases}\\
     &\label{eq:g2-ef-sols} XP_3 =
     \begin{cases}
         \left(456\right) \cap \left(X67\right) \qquad \an{X456} = 0\,,\\
         \left(X45\right) \cap \left(567\right) \qquad \an{X567} = 0\,.
    \end{cases}
\end{align}
The complete list of solutions, including two special configurations, are tabulated in Fig.~\ref{fig:graph2}. These special configurations arise because solutions \#1 and \#4 break down when $X =4$ and $X=5$ respectively. Analyzing (\ref{eq:g2-ll}-\ref{eq:g2-ef}) more carefully in these cases leads to solutions \#5 and \#6.

\begin{figure}
\centering
  \begin{minipage}[b]{0.45\textwidth}
      \centering
    \input{graph2.tex}
  \end{minipage}
  \hfill
  \begin{minipage}[c]{0.9\textwidth}
    \centering
     \begin{tabular}{|l|l|L| L| L|L|}\label{tabl:g2-o-sols}
        \# & $X$ is a point on the line & AB & CD & EF & \text{Letter classes}\\ \hline \hline
        1 & (234) $\cap$ (456), $\left(X \neq 4\right)$ & (1,X) & (4,X) & (6,X) & b_0, b_2, b_3, b_5\\
        2 & (234) $\cap$ (567) & (1, X) & (4,X) & (5,X) & b_0, b_2, b_3, b_4, c_1\\
        3 & (45) & (1, X) & (3,X) & (6,X)  & b_0, b_1, b_2, b_3, b_4, c_1\\
        4 & (345) $\cap$ (567), $\left(X \neq 5\right)$ & (1, X) & (3, X)& (5,X) &  b_0, b_3, b_5\\
        5 &  (23) & (1,4) & (4,X) & (4,6) & b_0, b_2, b_4\\
        6 &  (67) & (1,5) & (5,3) & (5,X) & b_0, b_3, b_5\\ \hline
         \end{tabular}
    \end{minipage}
    \caption{Cuts and first-type Landau singularities, up to parity conjugates, for the graph shown.}
    \label{fig:graph2}
  \end{figure}

Due to the presence of a degree of freedom in the on-shell solutions, solving the Kirchhoff equations requires adopting the strategy outlined in Section~3.2 of~\cite{Lippstreu:2022bib} and we find non-trivial solutions at specific values of the free parameter, some of which lead to letters outside the heptagon alphabet. These letters arise from cuts \#2 and \#3 (and their parity conjugates).

We will present the details of how the new letters arise from cut \#3. The on-shell solution
\begin{gather}
AB=(1,X)\,, \qquad CD=(3,X)\,, \qquad EF = (6,X)\,,
\end{gather}
where $X$ is an arbitrary point on the line $(4,5)$, determines the momenta $\ell_1, \ell_2$ flowing through the lines $(1,7)$ and $(1,2)$ to be
\begin{gather}
    \ell_{1}=\frac{\e(\cdot,1,7,X)}{\braket{17}\braket{1X}}Z_1\,,
    \qquad \ell_2=\frac{\e(\cdot,2,1,X)}{\braket{21}\braket{1X}}Z_1\,.
\end{gather}
A particular solution exists for $\alpha_3 = \alpha_4 = \dots = \alpha_8 = 0$, in which case the only Kirchhoff conditions that remain are
\begin{gather}
    \label{eq:g2-k1}
    \alpha_1 \, \ell_1 + \alpha_2 \, \ell_2 + \alpha_9 \,\ell_{\text{\tiny{ABEF}}} + \alpha_{10}\, \ell_{\text{\tiny{ABCD}}} = 0
\end{gather}
and
\begin{equation}
\begin{split}
\label{eq:g2-k2} \alpha_{11} \, \ell_{\text{\tiny{CDEF}}} + \alpha_{10}\, \ell_{\text{\tiny{ABCD}}}  &= 0\,,\\
 \alpha_{11} \, \ell_{\text{\tiny{CDEF}}} - \alpha_9 \, \ell_{\text{\tiny{ABEF}}} &= 0\,.
\end{split}
\end{equation}
The solution of interest is achieved by demanding
\begin{align}
    \label{eq:g2-k1-1}
    & \alpha_1 \, \ell_1 + \alpha_2 \, \ell_2 = 0
\end{align}
which requires imposing
\begin{gather}
    \braket{127X}= 0 \implies X =(45)\cap(127)\,, \label{eq:coll}
\end{gather}
fixing the value of the free parameter in $X$. On support of this solution, we have
\begin{gather}
   \ell_{\text{\tiny{ABCD}}} =\frac{\e(\cdot,1,l_{45},3)}{\braket{AB}\braket{CD}} X\,,
   \quad  \ell_{\text{\tiny{ABEF}}} =\frac{\e(\cdot,1,l_{45},6)}{\braket{AB}\braket{EF}} X\,,
   \quad \ell_{\text{\tiny{CDEF}}} =\frac{\e(\cdot,3,l_{45},6)}{\braket{CD}\braket{EF}} X\,,
\end{gather}
and~(\ref{eq:g2-k1}) now reads
\begin{gather}
\label{eq:g2-k1-2}
    \alpha_9 \,\ell_{\text{\tiny{ABEF}}} + \alpha_{10}\, \ell_{\text{\tiny{ABCD}}} = 0\,.
\end{gather}
A non-trivial solution to~(\ref{eq:g2-k2}, \ref{eq:g2-k1-2}) exists only when these three momenta are all collinear, which imposes
\begin{gather}
\label{eq:g2-letter1}
    \braket{136X}= \braket{136,(45)\cap(127)} =   \braket{1(27)(36)(45)} = 0\,,
\end{gather}
where we have used~(\ref{eq:coll}). We see the appearance of a singularity at $\braket{1(27)(36)(45)}=0$ which is outside the heptagon alphabet. Accounting for cyclic permutations of the graph and parity conjugate solutions results in two new classes of letters containing representatives
\begin{align}
    \label{eq:classc1}
    c_1 \ni \an{1(27)(36)(45)}\,, \qquad \bar{c}_1 \ni \an{1452}\an{1567}\an{7234}+ \an{1457}\an{1234}\an{2567}\,.
\end{align}
It is interesting to note that all of the new letters, both in this graph and in graph 1, arise from the special $Y$ configurations described in Section~\ref{sec:YDelta} that involve three collinear momenta at a $Y$ vertex.

\section{\texorpdfstring{Scalar versus $\mathcal{N}=4$ leading singularities}{Scalar versus N=4 leading singularities}}
\label{sec:five}

The Landau equations predict the singular loci of individual scalar Feynman integrals, but some singularities can certainly disappear if an integral is dressed with a non-trivial numerator factor. In this section, we probe this phenomenon in $\mathcal{N}=4$ SYM theory by comparing (for each graph, 1 and 2) a particular leading singularity of the \emph{scalar} integral associated to the graph and the corresponding $\mathcal{N}=4$ leading singularity computed via its \emph{on-shell diagram} (see~\cite{Arkani-Hamed:2012zlh}). In particular, we consider the cuts associated to singularities outside the heptagon alphabet. We find that these leading singularities in $\mathcal{N}=4$ SYM are free of letters outside the heptagon alphabet. These calculations thus serve as suggestive evidence that letters outside the heptagon alphabet are absent from the singularity structure of $\mathcal{N}=4$ SYM. The on-shell diagram associated to a specific cut (i.e., to a specific solution of the on-shell equations) is determined by the following procedure:
\begin{enumerate}
    \item Determine the momentum carried by each line in the graph by using the associated twistors and~(\ref{eq:tvmap}).
    \item The on-shell conditions ensure that this momentum is null and only one of the terms in (\ref{eq:tvmap}) survives. The $\tilde{\lambda}$ spinors are associated to the factor proportional to the $\epsilon$ tensor and the $\lambda$ spinors to the rest.
    \item Color a 3-point vertex black if the three momenta associated to it (call them $p_1, p_2, p_3$) are such that $\tilde{\lambda}_1 \propto \tilde{\lambda}_2 \propto \tilde{\lambda}_3$; color it white otherwise.
    \item Blow up every 4-point vertex into its on-shell diagram; there is a unique coloring of this diagram up to square moves.
\end{enumerate}
We freely utilize the machinery of on-shell diagrams and the ``positroids" Mathematica package~\cite{Bourjaily:2012gy} in the remainder of this section. We refer the reader to \cite{Arkani-Hamed:2012zlh} for more information about the technical details involved in these types of computations.

\subsection{Graph 1: Scalar leading singularity}

We compute the leading singularity on the solution
\begin{equation}
    \label{eq:ls1-sol}
\begin{aligned}
AB = \left(6, 671 \cap 34\right),\qquad
CD = \left(2,671 \cap 34\right),\qquad
EF = \left(5, 671 \cap 34\right),
\end{aligned}
\end{equation}
which is the parity conjugate of solution \#4 in Fig.~\ref{fig:graph1}. It is given by
\begin{multline}
    \label{eq:g1-scalarint}
    LS_1^{\text{scalar}} = {\oint} \frac{\an{ABd^2A}\an{ABd^2B}\an{CDd^2C}\an{CDd^2D}\an{EFd^2E}\an{EFd^2F}}{\an{AB67}\an{AB17}\an{AB56}\an{CD12}\an{CD23}\an{CD34}\an{EF34}\an{EF45}\an{EF56}}\\
 \times \frac{1}{\an{ABCD}\an{EFCD}\an{ABEF}}\,.
\end{multline}
By parameterizing
\begin{align}
    C = A + \alpha_1 B + \gamma_1 Z_{\star}\,, \qquad E = A + \alpha_2 B + \gamma_2 Z_{\star}
\end{align}
in terms of an arbitrary reference spinor $Z_{\star}$, we can easily compute the residue on
\begin{align}
    \an{ABCD}\an{ABEF}\an{EFCD} = \gamma_1 \gamma_2 \left(\alpha_1 - \alpha_2 \right) \an{AB\star D}\an{AB \star F} \an{ABDF} = 0\,.
\end{align}
Now setting $X = A + \alpha_1 B $ and  $\an{X d^3X} = \an{ABd^2A} d\alpha_1 $ we are left with
\begin{multline}
    LS_1^{\text{scalar}} = \oint \frac{\an{Xd^3X} \an{XBd^2B}}{\an{XBDF}\an{XB17}\an{XB67}\an{XB56}}\\
    \times \oint \frac{\an{XD d^2D} \an{XF d^2F}}{\an{XD12}\an{XD23}\an{XD34}\an{XF34}\an{XF45}\an{XF56}}\,.
\end{multline}
The integrals over $B, D$ and $F$ evaluate (keeping in mind that we intend to localize onto the solution (\ref{eq:ls2-sol})) to
\begin{align}
    \nonumber\oint \frac{\an{XBd^2B}}{\an{XB17}\an{XB67}\an{XB56}} &= \frac{1}{\an{\left(X 17\cap X 67\right)56}} \qquad \text{while setting }  B = X17 \cap 67\,,\\
    \oint \frac{\an{XDd^2D}}{\an{XD12}\an{XD23}\an{XD34}} &= \frac{1}{\an{\left(X 12\cap X 23\right)34}} \qquad \text{while setting }  D = X12 \cap 23\,,\\
    \nonumber\oint \frac{\an{XFd^2F}}{\an{XF34}\an{XF45}\an{XF56}} &= \frac{1}{\an{\left(X 34\cap X 45\right)56}} \qquad \text{while setting }  F = X34 \cap 56\,.
\end{align}
We are finally left with an integral over $X$:
\begin{align}
    LS_1^{\text{scalar}} &= \oint \frac{\an{Xd^3X}}{\an{X671}\an{X234}\an{X345}\an{X567}\an{X123}\an{X456}\an{X256}}\\
    & = \frac{1}{\an{3467}\an{1567}\an{1367}\an{1234}\an{1467}\an{3456}\an{6(17)(34)(25)}}\,.
\label{eq:ls1scalar}
\end{align}
In the last line we have computed the residue on $\an{X671} = \an{X234} = \an{X345} = 0$ which sets $X = (671) \cap (34)$. In~(\ref{eq:ls1scalar}) we see that this leading singularity has a pole at $\an{6(17)(34)(25)} = 0$. This singularity is outside the heptagon alphabet and is precisely the new letter found in~(\ref{eq:g2-letter1}).

\subsection{\texorpdfstring{Graph 1: $\mathcal{N}=4$ leading singularity}{Graph 1: N=4 leading singularity}}

\input{plabicgraphs.tex}

The on-shell diagram corresponding to the solution~(\ref{eq:ls1-sol}) is shown in the left panel of Fig.~\ref{fig:plabicgraphs}. It corresponds to the decorated permutation\footnote{This is the permutation labeling the corresponding cell in the momentum twistor Grassmannian.} $\left\lbrace 3,5,4,6,7,9,8\right\rbrace$ and the matrix
\begin{align}
\label{eq:plabicgraph1-cmatrix}
    C_1 = \begin{pmatrix}
        1 &\alpha_3 + \alpha_5 + \alpha_8  & \left(\alpha_3 + \alpha_5\right) \alpha_7 & \left(\alpha_3 + \alpha_5 \right)\alpha_6 & \alpha_3 \alpha_4 & 0 & -\alpha_1 \\
        0 & 1 & \alpha_7 & \alpha_6 & \alpha_4 & \alpha_2 & 0
    \end{pmatrix}.
\end{align}
The leading singularity is then computed as
\begin{align}
    LS_1^{\mathcal{N}=4} = \oint \prod_{i=1}^8 \frac{d\alpha_i}{\alpha_i} \delta^8 \left( C_1 \cdot Z\right).
\end{align}
(Upon introducing an appropriate fermionic delta function we identify this with the $\text{N}^2\text{MHV}$ Yangian invariant $[2,3,4,5,6][7,1,2,(234)\cap(56),6]$.) Integrating out the delta functions $\delta \left( C_1 \cdot Z \right)$ sets
\begin{equation}
\begin{split}
\label{eq:pg1sols}
    &\alpha_1 = \frac{\langle 1256\rangle }{\langle 2567\rangle }\,, \quad \alpha_2= \frac{\langle
   2345\rangle }{\langle 3456\rangle }\,, \quad\alpha_3 = \frac{\langle 3456\rangle  \langle 2(17)(34)(56)\rangle
   }{\langle 2345\rangle  \langle 2346\rangle  \langle 2567\rangle }\,, \quad\alpha_4 = \frac{\langle
   2346\rangle }{\langle 3456\rangle }\,,\\
   &\alpha_5 = \frac{\langle 1267\rangle  \langle 3456\rangle
   }{\langle 2346\rangle  \langle 2567\rangle }\,, \quad\alpha_6 = \frac{\langle 2356\rangle }{\langle
   3456\rangle }\,,\quad \alpha_7 = \frac{\langle 2456\rangle }{\langle 3456\rangle }\,,\quad \alpha_8 = \frac{\langle
   1567\rangle }{\langle 2567\rangle}
\end{split}
\end{equation}
and introduces a Jacobian factor
\begin{align}
    \mathcal{J}_1 = \frac{\an{2345}\an{2346}\an{2567}}{\an{3456}}\,.
\end{align}
Putting everything together, we arrive at
\begin{align}
\label{eq:ls1n=4}
    LS_1^{\mathcal{N}=4} = \frac{\an{2567}^3 \an{3456}^3}{\an{1256}\an{1267}\an{1567}\an{2345}\an{2356}\an{2456}\an{2(17)(34)(56)}}\,.
\end{align}
The non-heptagon letter $\an{6(17)(34)(25)}$ appearing in the scalar leading singularity~(\ref{eq:ls1scalar}) is absent in~(\ref{eq:ls1n=4}), indicating cancellations occurring in $\mathcal{N}=4$ SYM theory.

\subsection{Graph 2: Scalar leading singularity}

For this graph, we choose to compute the residue on solution \#3 in Fig.~\ref{fig:graph2}:
\begin{align}
\label{eq:ls2-sol}
AB = (1,X)\,, \qquad CD = (3,X)\,, \qquad EF=(6,X)\,,
\end{align}
with $X$ an arbitrary point on the line $(45)$. The leading singularity of the scalar integral is
\begin{multline}
    \label{eq:g2-scalarint}
    LS^{\text{scalar}}_2 = {\oint} \frac{\an{ABd^2A}\an{ABd^2B}\an{CDd^2C}\an{CDd^2D}\an{EFd^2E}\an{EFd^2F}}{\an{AB12}\an{AB17}\an{CD23}\an{CD34}\an{CD45}\an{EF45}\an{EF56}\an{EF67}}\\
     \times \frac{1}{\an{ABCD}\an{EFCD}\an{ABEF}}\,.
\end{multline}
A convenient parameterization is
\begin{align}
    \label{eq:g2-residue-par}
    &\nonumber A = Z_1 + x_1 Z_{\star_1}+ x_2 Z_{\star_2}\,, \qquad B = Z_4 + y_1 Z_5 + x_3 Z_{\star_3}\,, \\
    & C = Z_3 + x_4 Z_{\star_4}+ x_5 Z_{\star_5}\,, \qquad D = Z_4 + y_2 Z_5 + x_6 Z_{\star_6}\,, \\
    &\nonumber E = Z_6 + x_7 Z_{\star_7}+ x_8 Z_{\star_8}\,, \qquad F = Z_5 + y_3 Z_4 + x_9 Z_{\star_9}\,,
\end{align}
where $Z_{\star_i}$ are arbitrary twistors. The integral over $E$ is then
\begin{align}
    &\nonumber\oint \frac{\an{EFd^2E}}{\an{EF56}\an{EF67}}\\
    &\qquad= \underset{x_8 = 0}{\oint} dx_8 \underset{x_7 = -\frac{\an{\star_8 F56}x_8}{\an{\star_7 F56}}}{\oint} dx_7 \frac{\an{6F\star_7 \star_8}}{\left(\an{\star_7 F56} x_7 + \an{\star_8 F56} x_8\right)\left(\an{\star_7 F67} x_7 + \an{\star_8 F67} x_8\right)}\\
    &\qquad\nonumber=\frac{1}{\an{F567}}\,.
\end{align}
A similar contour for $x_9$ lets us carry out part of the $F$ integral as
\begin{align}
    \oint \frac{\an{EFd^2F}}{\an{F567}\an{EF45}} =\frac{1}{\an{4567}}\int \frac{dy_3}{y_3}\,,
\end{align}
with the contour for the $y_3$ integral to be specified later. Performing the $C, D$ and $A$ integrals then gives
\begin{align}
     &\oint \frac{\an{CDd^2C}\an{CDd^2D}}{\an{CD23}\an{CD34}\an{CD45}} = \frac{1}{\an{2345}}\int \frac{dy_2}{y_2}\,, \qquad \oint \frac{\an{ABd^2A}}{\an{AB12}\an{AB17}} = \frac{1}{\an{B172}}\,.
\end{align}
Finally, picking contours for $y_1, y_2$ and $x_3$ which compute the residues on the poles $\an{ABCD} = \an{ABEF} = \an{CDEF}= 0$ leaves us with an integral over the residual degree of freedom $y_1$:
\begin{align}
\label{eq:ls2scalar}
    LS^{\text{scalar}}_2  = \frac{1}{\an{4567}\an{2345}\an{3456}} \int \frac{dy_1}{\left(\an{4127}+y_1 \an{5127}\right)\left(\an{1436}+y_1\an{1536}\right)}\,.
\end{align}
This integral has a singularity when the two poles collide, which occurs when
\begin{align}
    \an{1536}\an{1427}-\an{1436}\an{1527} = \an{1(27)(36)(45)}=0
\end{align}
precisely in agreement with the letter found in~(\ref{eq:g2-letter1}).

\subsection{\texorpdfstring{Graph 2: $\mathcal{N}=4$ leading singularity}{Graph 1: N=4 leading singularity}}

The on-shell diagram corresponding to the solution~(\ref{eq:ls2-sol}) is shown in the right panel of Fig.~\ref{fig:plabicgraphs}. It is associated to the permutation $\left\lbrace 3,4,6,5,7,1,2\right\rbrace$ and the matrix
\begin{align}
\label{eq:plabicgraph2-cmatrix}
  C_2 =  \left(
\begin{array}{ccccccc}
 1 & \alpha_{2}+\alpha_{4}+\alpha_{7}+\alpha_{9} & (\alpha_{2}+\alpha_{4}+\alpha_{7}) \alpha_{8} & (\alpha_{2}+\alpha_{4}) \alpha_{6} & (\alpha_{2}+\alpha_{4}) \alpha_{5} & \alpha_{2} \alpha_{3} & 0 \\
 0 & 1 & \alpha_{8} & \alpha_{6} & \alpha_{5} & \alpha_{3} & \alpha_{1} \\
\end{array}
\right).
\end{align}
The leading singularity computed via this on-shell diagram is
\begin{align}
    LS_2^{\mathcal{N}=4} &= \oint \prod_{i=1}^9 \frac{d\alpha_i}{\alpha_i} \delta^8 \left( C_2 \cdot Z\right).
\end{align}
The constraints from $\delta\left(C_2 \cdot Z \right)$ fix $8$ of the $9$ $\alpha_i$ to be
\begin{equation}
\begin{split}
    &\alpha_1 = \frac{\an{6(17)(23)(45)+\alpha_9 \an{2367}\an{2456}}}{\an{4567}\left(\an{1367}+\alpha_9 \an{2367}\right)}\,, \\
     &\alpha_2 = -\frac{\an{4567}\left(\an{1367}+\alpha_9 \an{2367}\right)\left(\an{1(23)(45)(67)-\an{2(13)(45)(67)\alpha_9}}\right)}{\left(\an{2367}\an{2457}\alpha_9 -\an{7(16)(23)(45)}\right) \left( \an{6(17)(23)(45)}+\alpha_9 \an{2367}\an{2456} \right)}\,,\\
    & \alpha_3 = \frac{\langle 7(16)(23)(45)\rangle -\alpha _{9} \langle 2367\rangle  \langle 2457\rangle }{\langle 4567\rangle (\alpha _{9} \langle 2367\rangle +\langle 1367\rangle )}\,,\\
    &\alpha_4 = \frac{\langle 1237\rangle  \langle 4567\rangle  (\alpha _{9} \langle 2367\rangle +\langle 1367\rangle)}{\langle 2367\rangle  (\langle 7(16)(23)(45)\rangle -\alpha _{9} \langle 2367\rangle  \langle 2457\rangle
   )}\,,\\
   & \alpha_5 = \frac{\langle 2367\rangle  (\alpha _{9} \langle 2467\rangle +\langle 1467\rangle )}{\langle 4567\rangle (\alpha _{9} \langle 2367\rangle +\langle 1367\rangle )}\,, \qquad \alpha_6 =  -\frac{\langle 2367\rangle  (\alpha _{9} \langle 2567\rangle +\langle 1567\rangle )}{\langle 4567\rangle (\alpha _{9} \langle 2367\rangle +\langle 1367\rangle )}\,,\\
   & \alpha_7 = -\frac{\alpha _{9} \langle 2367\rangle +\langle 1367\rangle }{\langle 2367\rangle}\,, \qquad \alpha_8 = -\frac{\langle 1267\rangle }{\alpha _{9} \langle 2367\rangle +\langle 1367\rangle}
\end{split}
\end{equation}
and introduce the Jacobian factor
\begin{align}
    \mathcal{J}_2 =- \frac{(\alpha_{9} \langle 2367\rangle  \langle 2456\rangle +\langle 6(17)(23)(45)\rangle ) (\langle
   7(16)(23)(45)\rangle -\alpha_{9} \langle 2367\rangle  \langle 2457\rangle )}{\langle 4567\rangle  (\alpha_{9} \langle 2367\rangle +\langle 1367\rangle )}\,.
\end{align}
Altogether we are left with a one-parameter integral, as expected since $C_2$ is a 9-dimensional cell of the Grassmannian ${\rm Gr}\left(2,7\right)$:
\begin{multline}
\label{eq:ls2n=4}
    LS_2^{\mathcal{N}=4}
    = \frac{\langle 4567\rangle ^3}{\langle 1237\rangle\,
    \langle 1267\rangle}\int \frac{d\alpha_9}{\alpha_9} \, \frac{(\alpha_{9} \langle 2367\rangle +\langle 1367\rangle )^3}{\left(\alpha_{9} \langle 2367\rangle  \langle 2456\rangle +\langle 6(17)(23)(45)\rangle \right) \, \left(\alpha_{9} \langle 2467\rangle +\langle 1467\rangle \right)}\\
      \times\frac{1}{\left(\alpha_{9} \langle 2567\rangle +\langle
   1567\rangle \right)\, \left(\alpha_{9} \langle 2(13)(45)(67)\rangle -\langle 1(23)(45)(67)\rangle \right)}\,.
\end{multline}
This integral has a singularity when any two of its four poles collide. A quick calculation reveals that the singular locus is
\begin{multline}
\an{1235}\an{1236}\an{1237}\an{1267}\an{2345}
\an{2346}\an{2356}\an{2456}\an{2467}\an{2567}\\
\times \an{6(17)(23)(45)}\an{2(13)(45)(67)} = 0\,.
\end{multline}
(This can also be checked by computing all residues and tabulating their poles). Remarkably, this is free of non-heptagon symbol letters, again exhibiting special cancellations in $\mathcal{N}=4$ SYM theory.

\section{Discussion}
\label{sec:discussion}

We have solved the Landau equations for nine three-loop 7-point graphs, two in Section~\ref{sec:newletters} and seven more in Appendix~\ref{sec:othergraphs}. The singularities of the latter are all consistent with the heptagon symbol alphabet, whereas the former have additional singularities at the vanishing locus of 28 new letters (equations~(\ref{eq:classc0}), (\ref{eq:classc1}) and their cyclic images). If it is somewhat surprising that there are massless planar 7-point graphs with singularities outside those of the heptagon alphabet, it is certainly more surprising that we did not encounter a vast zoo of new singularities but only a rather tame collection of new characters.

In particular, we note that all of the new letters are non-dihedral permutations\footnote{The curious fact that the hexagon symbol alphabet is invariant under arbitrary permutations of external legs, but the heptagon alphabet is not, was noted in~\cite{Drummond:2014ffa}. The fact that we encounter non-dihedral permutations suggests something with a ``non-planar'' flavor hiding inside a planar diagram. A phenomenon like this already occurs for the massless box integral, which has a branch point at $(p_1 + p_3)^2 = 0$ beginning at ${\cal{O}}(\epsilon)$ in dimensional regularization (we thank A.~McLeod for this comment). That case, however, arises from a second-type Landau singularity whereas we consider only first-type singularities in this paper.} of heptagon letters (or parity conjugates thereof). However, we only encountered a few out of all possible non-dihedral permutations, and it would be interesting to investigate whether others are realized by more complicated higher-loop graphs, or if there is some not yet understood magic that makes certain permutations special.

It is also interesting to note that half of the new letters (those in~(\ref{eq:classc0})) are precisely the ones predicted by analyzing on-shell diagrams corresponding to Yangian invariants~\cite{He:2020uhb} and Schubert problems~\cite{Yang:2022gko}, while the other half do not appear in those analyses\footnote{However, we thank Qinglin Yang for pointing out that it is possible to formulate a Schubert problem in which they do appear.}. It is also worth noting that none of the new letters have a fixed sign for positive external kinematics ($\an{ijkl}>0$ for $i<j<k<l$); therefore they are not cluster variables of ${\rm Gr}(4,7)$.

The relevance of the new symbol letters for $\mathcal{N}=4$ SYM theory remains unclear, though the analysis in Section~\ref{sec:five}---not to mention the success of the heptagon bootstrap to impressively high loop order~\cite{Dixon:2013eka,Dixon:2014xca,Caron-Huot:2016owq,Caron-Huot:2019vjl,Dixon:2021nzr,Dixon:2020cnr}, evidently without the need for any of these new letters---suggests they may be absent to all loop order. A formula for the integrand of the three-loop 7-point MHV amplitude in $\mathcal{N}=4$ SYM theory was presented in~\cite{Arkani-Hamed:2010pyv}, and it includes contributions from integrals having the same underlying scalar topology as our graphs 1 and 2. These individual scalar integrals have singularities absent from the full amplitude due to some combination of (1) possible cancellation among various terms in the sum and (2) possible cancellation of certain singularities due to the very special numerator factors appearing in the integrals of~\cite{Arkani-Hamed:2010pyv}. We have checked that the numerator factors attached to graphs 1 and 2 vanish on all solutions of the Landau equations that exhibit a new letter. We would like to emphasize that these calculations are merely suggestive and do not constitute a proof of the completeness of the heptagon alphabet. It would be interesting to develop a more rigorous argument which holds at all loop orders.

In the course of this work, we discovered that there exist solutions of the Landau equations that evade the $Y{-}\Delta$ equivalence used in the analysis of~\cite{Prlina:2018ukf} to reduce arbitrary massless planar graphs to ``ziggurat'' graphs. Therefore, the interesting problem of determining the locus of possible singularities of these graphs in general theories, and the more limited but likely more tractable problem of proving (or disproving) the conjectured completeness of the 6- and 7-point symbol alphabets for amplitudes in $\mathcal{N}=4$ SYM theory, remain open.

\acknowledgments

We are grateful to Nima Arkani-Hamed, Lance Dixon, Hofie Hannesdottir, Andrew McLeod, Sebastian Mizera, Andrzej Pokraka, Lauren Williams and Qinglin Yang for helpful discussions. This work was supported in part by the US Department of Energy under contract {DE}-{SC}0010010 Task F and by Simons Investigator Award \#376208.

\appendix

\section{Landau singularities of various three-loop 7-point graphs}
\label{sec:othergraphs}

In this Appendix, we present all solutions to the on-shell conditions associated to the seven three-loop graphs shown in Figures~\ref{fig:firstfig}--\ref{fig:lastfig}, and the types of singularities they possess as determined by solving the corresponding Kirchhoff equations. The first six graphs each have eight cuts, while the last has twelve. In the first three columns of the tables below, we indicate half of the solutions to each set of on-shell conditions; the other half are their parity conjugates. In the tables, we use $l_{AB}(z) = z A + (1 - z) B$ to parameterize an arbitrary point on the line containing $A$ and $B$. The final column of each table indicates the ``letter classes'' associated to each cut, meaning that the Kirchhoff conditions can be satisfied only when the external kinematics are such that a heptagon symbol letter of one of the indicated types (using the notation of~(\ref{eq:heptagonalphabet})) vanishes. All seven graphs have singularities corresponding only to the heptagon alphabet.

\begin{figure}[htb!]
\centering
  \begin{minipage}[c]{0.35\textwidth}
      \centering
    \input{graph3.tex}
  \end{minipage}
  \hfill
  \begin{minipage}[c]{0.6\textwidth}
    \centering
     \begin{scriptsize}
     \begin{tabular}{|l|L| L| L| L|}
         \# & AB & CD & EF & \text{Letter classes}  \\ \hline \hline
         1 & (7,5) & (5,3) & (1,3) & b_4 \\
         2 & (356)\cap\bar{7} & (5,3) & (1,3)  & b_4,b_2\\
         3 & (7,56\cap 134) & (4,\bar{5}\cap 13) & (1,3) & b_4,b_1  \\
         4 & \bar{5}\cap\bar{7}& (4,\bar{5}\cap 13) & (1,3) & b_2 \\ \hline
         \end{tabular}
          \end{scriptsize}
    \end{minipage}
    \caption{Cuts and first-type Landau singularities, up to parity conjugates, for the graph shown.}
 \label{fig:firstfig}
\end{figure}

\begin{figure}[htb!]
\centering
  \begin{minipage}[c]{0.35\textwidth}
      \centering
    \input{graph4.tex}
  \end{minipage}
  \hfill
  \begin{minipage}[c]{0.6\textwidth}
    \centering
     \begin{scriptsize}
      \begin{tabular}{|l|L| L| L| L|}
        \# & AB & CD & EF& \text{Letter classes}  \\ \hline \hline
        1 & (1,3) & (6,1) & (4,6) & b_0,b_2\\
        2 & (1,3) & (6,1) & (5, \bar{4}\cap 61)  & b_0,b_2 \\
        3 & (1,3) & (7,\bar{6}\cap 13) & (4,56\cap 713) & b_0,b_2,b_4   \\
        4 & (1,3) & (7,\bar{6}\cap 13) & (5,34\cap \bar{6}) &  b_0,b_2,b_4 \\ \hline
         \end{tabular}
          \end{scriptsize}
              \end{minipage}
     \caption{Cuts and first-type Landau singularities, up to parity conjugates, for the graph shown.}
     \label{fig:graph4}
  \end{figure}

\begin{figure}[htb!]
\centering
  \begin{minipage}[c]{0.35\textwidth}
      \centering
    \input{graph5.tex}
  \end{minipage}
  \hfill
  \begin{minipage}[c]{0.6\textwidth}
    \centering
     \begin{scriptsize}
     \begin{tabular}{|l|L| L| L| L|}
       \# & AB & CD & EF & \text{Letter classes}  \\ \hline \hline
        1 &  (2,6) & (2,4) & (4,6) & b_0,b_2,b_3,b_4\\
        2 &  (267)\cap(245) & (2,4) & 245\cap \bar{6}  & b_0,b_2,b_3,b_4,b_6\\
        3 &  (2,6) & \bar{4}\cap (236) & \bar{5}\cap (236) & b_0,b_1,b_2,b_3  \\
        4 &  (2,67\cap \bar{4}) & (367)\cap \bar{4} & \bar{6}\cap\bar{4} & b_0,b_1,b_2,b_3,b_5 \\ \hline
         \end{tabular}
          \end{scriptsize}
              \end{minipage}
      \caption{Cuts and first-type Landau singularities, up to parity conjugates, for the graph shown.}
      \label{fig:graph5}
  \end{figure}

\newpage
\begin{figure}[htb!]
\centering
  \begin{minipage}[c]{0.35\textwidth}
      \centering
    \input{graph6.tex}
  \end{minipage}
  \hfill
  \begin{minipage}[c]{0.6\textwidth}
    \centering
     \begin{scriptsize}
     \begin{tabular}{|l|L| L| L| L|}
       \# & AB & CD & EF & \text{Letter classes}  \\ \hline \hline
        1 &  (1,3) & \bar{6}\cap (713) & (4,13\cap \bar{6}) & b_{0},b_1,b_2,b_3 \\
        2 &  (1,3) & \bar{6}\cap (713) & (3,\bar{4}\cap 17)  & b_0,b_1,b_2,b_3 \\
        3 & (1,3) & (6,1) & (4,1) & b_0,b_2,b_3,b_4   \\
        4 &  (1,3) & (6,1) & (3,16\cap \bar{4}) &  b_0,b_2,b_3,b_4  \\ \hline
         \end{tabular}
          \end{scriptsize}
              \end{minipage}
   \caption{Cuts and first-type Landau singularities, up to parity conjugates, for the graph shown.}
   \label{fig:graph6}
  \end{figure}

\begin{figure}[htb!]
\centering
  \begin{minipage}[c]{0.35\textwidth}
      \centering
    \input{graph7.tex}
  \end{minipage}
  \hfill
  \begin{minipage}[c]{0.6\textwidth}
    \centering
     \begin{scriptsize}
    \begin{tabular}{|l|L| L| L| L|}
         \# & AB & CD & EF & \text{Letter classes}  \\ \hline \hline
         1 & (1,l_{23}(x)) & (4AB)\cap (4EF) & (6,l_{17}(y)) & b_0,b_1,b_3,b_6 \\
         2 &  (1,l_{23}(x)) & (4AB)\cap (4EF) & (7,l_{56}(y))  & b_0,b_1,b_6 \\
         3 &  (2,l_{17}(x)) & (4AB)\cap (4EF) & (6,l_{17}(y)) & b_0,b_1,b_6  \\
         4 &  (2,l_{17}(x)) & (4AB)\cap (4EF) & (7,l_{56}(y)) &  b_0,b_1,b_2,b_6\\ \hline
         \end{tabular}
          \end{scriptsize}
              \end{minipage}
  \caption{Cuts and first-type Landau singularities, up to parity conjugates, for the graph shown.}
  \label{fig:graph7}
  \end{figure}

\begin{figure}[htb!]
\centering
  \begin{minipage}[c]{0.35\textwidth}
      \centering
    \input{graph8.tex}
  \end{minipage}
  \hfill
  \begin{minipage}[c]{0.6\textwidth}
    \centering
     \begin{scriptsize}
    \begin{tabular}{|l|L| L| L| L|}
       \# & AB & CD & EF & \text{Letter classes}  \\ \hline \hline
        1 & (1,3) & (4,\bar{5}\cap (EF)) & (7,l_{AB}(x))& b_0,b_1,b_2\\
        2 & (1,3) & (5,\bar{4}\cap (EF))  & (7,l_{AB}(x))  & b_0,b_1,b_2 \\
        3 & \bar{1}\cap\bar{3} & (4,\bar{5}\cap (EF)) & (7,l_{AB}(x)) & b_0,b_1,b_3   \\
        4 &   \bar{1}\cap\bar{3} & (5,\bar{4}\cap (EF))& (7,l_{AB}(x)) &  b_0,b_1,b_3 \\ \hline
         \end{tabular}
          \end{scriptsize}
              \end{minipage}
    \caption{Cuts and first-type Landau singularities, up to parity conjugates, for the graph shown.}
  \end{figure}

\begin{figure}[htb!]
\centering
  \begin{minipage}[c]{0.35\textwidth}
      \centering
    \input{graph9.tex}
  \end{minipage}
  \hfill
  \begin{minipage}[c]{0.6\textwidth}
    \centering
     \begin{scriptsize}
    \begin{tabular}{|l|L| L| L| L|}
        \# & AB & CD & EF & \text{Letter classes}  \\ \hline \hline
         1 & (\bar{2}\cap (14l_{56}(x))) & (4,l_{56}(x)) & \bar{5}\cap \bar{7}& b_0,b_1  \\
         2 & (\bar{1}\cap (24l_{56}(x))) & (4,l_{56}(x)) & \bar{5}\cap\bar{7}  & b_0,b_1 \\
         3 & (\bar{2}\cap (14l_{56}(x))) & (4,l_{56}(x)) & (7,l_{56}(x)) &    b_0,b_1,b_4\\
         4 &  (\bar{1}\cap (24l_{56}(x))) & (4,l_{56}(x)) & (7,l_{56}(x)) & b_0,b_1,b_4 \\
         5 & (\bar{2}\cap (146)) & (4,6) & (6,l_{17}(x))& b_0,b_4\\
         6 & (\bar{1}\cap (246)) & (4,6) & (6,l_{17}(x))& b_0,b_4 \\ \hline
         \end{tabular}
          \end{scriptsize}
              \end{minipage}
 \caption{Cuts and first-type Landau singularities, up to parity conjugates, for the graph shown.}
   \label{fig:lastfig}
  \end{figure}

\end{document}

%% file: YDelta.tex
\begin{tikzpicture}
\filldraw[black] (2,2) circle (2pt) ;
\filldraw[black] (2,0) circle (2pt);
\filldraw[black] (4,4) circle (2pt);
\filldraw[black] (0,4) circle (2pt);
\node at (2,-0.5) {$p_1$};
\node at (4.5,4.5) {$p_2$};
\node at (-0.5,4.5) {$p_3$};
\draw[black, thick] (2,0) -- (2,1) node[anchor=south west]{\small{$\alpha_1$}};
\draw[<-, black, thick](2,1) -- (2,2) ;
\draw[->, black, thick] (2,2) -- (3,3)node[anchor=south east]{\small{$\alpha_2$}};
\draw[-, black, thick] (3,3) -- (4,4);
\draw[->, black, thick] (2,2) -- (1,3) node[anchor=south west]{\small{$\alpha_3$}};
\draw[black, thick] (1,3) -- (0,4);
\draw[red, thick, <-] (1.5,0.4) arc (0:60:3);
\draw[red, thick, ->] (2.5,0.4) arc (180:120:3);
\draw[red, thick, ->] (3.3,4.4) arc (340:200:1.4);
\node at (0.5,1.5) {$X_1$};
\node at (3.5,1.5) {$X_2$};
\node at (2,4) {$X_3$};
\begin{scope}[xshift=7cm]
\filldraw[black] (2,0) circle (2pt);
\filldraw[black] (4,4) circle (2pt);
\filldraw[black] (0,4) circle (2pt);
\node at (2,-0.5) {$p_1$};
\node at (4.5,4.5) {$p_2$};
\node at (-0.5,4.3) {$p_3$};
\draw[thick, -] (2.5,2) -- (3,3);
\draw[thick, ->] (2,1) -- (2.5,2);
\draw[thick, <-] (1.5,2) -- (1,3);
\draw[thick, -] (2,1) -- (1.5,2);
\draw[thick, -] (1,3) -- (2,3);
\draw[thick, <-] (2,3) -- (3,3);
\draw[->, black, thick] (2,1) -- (2,0.5);
\draw[-, black, thick] (2,0.5) -- (2,0);
\draw[->, black, thick] (3,3) -- (3.5,3.5);
\draw[-, black, thick] (3.5,3.5) -- (4,4);
\draw[->, black, thick] (1,3) -- (0.5,3.5);
\draw[-, black, thick] (0.5,3.5) -- (0,4);
\draw[red, thick, <-] (1.5,0.4) arc (0:60:3);
\draw[red, thick, ->] (2.5,0.4) arc (180:120:3);
\draw[red, thick, ->] (3.3,4.6) arc (340:200:1.4);
\node at (0.5,1.5) {$X_1$};
\node at (3.5,1.5) {$X_2$};
\node at (2,4) {$X_3$};
\node at (2,2.7) {\small{$\beta_1$}};
\node at (2,3.3) {\small{$q_2$}};
\node at (1.8,2.2) {\small{$\beta_2$}};
\node at (1.15,2.2) {\small{$q_3$}};
\node at (2.3,2.2) {\small{$\beta_3$}};
\node at (2.9,2.2) {\small{$q_1$}};
\end{scope}
\end{tikzpicture}

%% file: graph1.tex
\begin{tikzpicture}[line cap=round,line join=round,>=triangle 45,x=1cm,y=1cm, scale=0.8]
\draw [thick] (0,1)-- (1.02,1.73);
\draw [thick] (2,1)-- (1.02,1.73);
\draw [thick] (1.02,1.73)-- (1.02,2.81);
\node at (0,2.2) {\tiny{AB}};
\node at (1.4,2.2) {\tiny{$\alpha_{10}$}};
\node at (2,2.2) {\tiny{CD}};
\node at (1.3,1.2) {\tiny{$\alpha_{11}$}};
\node at (1,0.7) {\tiny{EF}};
\node at (0.4,1.65) {\tiny{$\alpha_{12}$}};
\draw [thick] (0,0)-- (-0.34,-0.97) node[below]{5};
\draw [thick] (-0.12,3.35)-- (-0.68,4.17)node[above]{7};
\draw [thick] (2.08,3.35)-- (2.36,4.19) node[above]{2};
\draw [thick] (3.24,1.85)-- (4.12,1.91)node[right]{3};
\draw [thick] (2,0)-- (2.38,-1.03)node[below]{4};
\draw [thick] (-1.2,1.95)-- (-2.14,2.13)node[left]{6}; 
\draw [thick] (1.02,2.81)-- (1.02,3.95)node[above]{1};
\draw [thick] (0,0)-- (2,0) ;
\draw [thick] (2,0)-- (2,1);
\draw [thick] (0,1)-- (0,0);
\draw [thick] (-0.12,3.35)-- (-1.2,1.95);
\draw [thick] (1.02,2.81)-- (-0.12,3.35);
\draw [thick] (-1.2,1.95)-- (0,1);
\draw [thick] (1.02,2.81)-- (2.08,3.35);
\draw [thick] (2.08,3.35)-- (3.24,1.85);
\draw [thick] (3.24,1.85)-- (2,1);
\node at (2.4, 0.5) {\tiny{$\alpha_5$}};
\node at (1,-0.4) {\tiny{$\alpha_6$}};
\node at (-0.5, 0.5) {\tiny{$\alpha_7$}};
\node at (-1, 1.2) {\tiny{$\alpha_8$}};
\node at (-1, 2.8) {\tiny{$\alpha_9$}};
\node at (0.5, 3.3) {\tiny{$\alpha_1$}};
\node at (1.4, 3.3) {\tiny{$\alpha_2$}};
\node at (3, 2.7) {\tiny{$\alpha_3$}};
\node at (2.9, 1.1) {\tiny{$\alpha_4$}};
\end{tikzpicture}

%% file: graph2.tex
\begin{tikzpicture}[line cap=round,line join=round,>=triangle 45,x=1cm,y=1cm, scale=0.7]
\draw [thick](0,1)-- (1,1.93);
\draw [thick](1,1.93)-- (1.02,2.81);
\draw [thick](1.02,2.81)-- (-0.12,3.35);
\draw [thick](-0.12,3.35)-- (-1.2,1.95);
\draw [thick](-1.2,1.95)-- (0,1);
\draw [thick](1.02,2.81)-- (2.08,3.35);
\draw [thick](2.08,3.35)-- (3.24,1.85);
\draw [thick](3.24,1.85)-- (2,1);
\draw [thick](2,1)-- (1,1.93);
\draw [thick](1,1.93)-- (1.02,2.81);
\draw [thick](-0.12,3.35)-- (-0.68,4.17)node[above]{4};
\draw [thick](2.08,3.35)-- (2.36,4.19)node[above]{5};
\draw [thick](3.24,1.85)-- (4.12,1.91)node[right]{6};
\draw [thick](-1.2,1.95)-- (-2.14,2.13)node[left]{3};
\draw [thick](0.98,0)-- (2,1);
\draw [thick](0,1)-- (0.98,0);
\draw [thick](0.98,0)-- (0.98,-1.09)node[below]{1};
\draw [thick](2,1)-- (2.86,-0.03)node[below]{7};
\draw [thick](0,1)-- (-0.88,0.01)node[below]{2};
\node at (0,2.5) {\tiny{CD}};
\node at (2,2) {\tiny{EF}};
\node at (1,0.9) {\tiny{AB}};
\node at (1.8,0.3) {\scriptsize{$\alpha_1$}};
\node at (0.3,0.3) {\scriptsize{$\alpha_2$}};
\node at (-0.8,1.3) {\scriptsize{$\alpha_3$}};
\node at (-1.0,2.9) {\scriptsize{$\alpha_4$}};
\node at (0.5,3.4) {\scriptsize{$\alpha_5$}};
\node at (1.5,3.4) {\scriptsize{$\alpha_6$}};
\node at (3,2.8) {\scriptsize{$\alpha_7$}};
\node at (2.8,1.2) {\scriptsize{$\alpha_8$}};
\node at (0.2,1.7) {\scriptsize{$\alpha_{10}$}};
\node at (1.5,2.5) {\scriptsize{$\alpha_{11}$}};
\node at (1.2,1.3) {\scriptsize{$\alpha_{9}$}};
\end{tikzpicture}

%% file: plabicgraphs.tex
\begin{figure}
\centering
\begin{tikzpicture}[line cap=round,line join=round,>=triangle 45,x=1cm,y=1cm]
\draw[thick] (3,4)-- (2,3);
\draw[thick] (2,3)-- (3,2);
\draw[thick] (3,2)-- (4,3);
\draw[thick] (4,3)-- (3,4);
\draw[thick] (3,2)-- (3,1);
\draw[thick] (3,1)-- (2,0);
\draw[thick] (2,0)-- (1,1);
\draw[thick] (1,1)-- (1,2);
\draw[thick] (1,2)-- (2,3);
\draw[thick] (2,3)-- (3,2);
\draw[thick] (4,3)-- (5,2);
\draw[thick] (5,2)-- (5,1);
\draw[thick] (5,1)-- (4,0);
\draw[thick] (4,0)-- (3,1);
\draw[thick] (3,1)-- (3,2);
\draw[thick] (3,2)-- (4,3);
\draw[thick] (2,0)-- (3,1);
\draw[thick] (3,1)-- (4,0);
\draw[thick] (4,0)-- (4,-1);
\draw[thick] (4,-1)-- (2,-1);
\draw[thick] (2,-1)-- (2,0);
\draw[thick] (3,4)-- (3,5)node[above]{1};
\draw[thick] (5,2)-- (5.78,2.53)node[right]{2};
\draw[thick] (5,1)-- (5.8,0.47)node[right]{3};
\draw[thick] (4,-1)-- (4.62,-1.63)node[below]{4};
\draw[thick] (2,-1)-- (1.46,-1.71)node[below]{5};
\draw[thick] (1,1)-- (0.14,0.41)node[left]{6};
\draw[thick] (1,2)-- (0.12,2.47)node[left]{7};
\begin{scriptsize}
\draw [fill=black] (3,4) circle (2.5pt);
\draw [fill=white] (2,3) circle (3pt);
\draw [fill=black] (3,2) circle (2.5pt);
\draw [fill=white] (4,3) circle (3pt);
\draw [fill=black] (3,1) circle (2.5pt);
\draw [fill=black] (2,0) circle (2.5pt);
\draw [fill=white] (1,1) circle (3pt);
\draw [fill=black] (1,2) circle (2.5pt);
\draw [fill=white] (5,2) circle (3pt);
\draw [fill=black] (5,1) circle (2.5pt);
\draw [fill=white] (4,0) circle (3pt);
\draw [fill=black] (4,-1) circle (2.5pt);
\draw [fill=white] (2,-1) circle (3pt);
\end{scriptsize}
\end{tikzpicture}
    \begin{tikzpicture}[line cap=round,line join=round,>=triangle 45,x=1cm,y=1cm]
\draw [thick] (3,2)-- (3,1);
\draw [thick] (3,1)-- (2,0);
\draw [thick] (2,0)-- (1,1);
\draw [thick] (1,1)-- (1,2);
\draw [thick] (1,2)-- (2,3);
\draw [thick] (2,3)-- (3,2);
\draw [thick] (4,3)-- (5,2);
\draw [thick] (5,2)-- (5,1);
\draw [thick] (5,1)-- (4,0);
\draw [thick] (4,0)-- (3,1);
\draw [thick] (3,1)-- (3,2);
\draw [thick] (3,2)-- (4,3);
\draw [thick] (5,2)-- (5.78,2.53) node[right]{6};
\draw [thick] (1,2)-- (0.12,2.47)node[left]{3};
\draw [thick] (2,0)-- (3,1);
\draw [thick] (3,1)-- (4,0);
\draw [thick] (4,0)-- (4,-1);
\draw [thick] (4,-1)-- (3,-2);
\draw [thick] (3,-2)-- (2,-1);
\draw [thick] (2,-1)-- (2,0);
\draw [thick] (1,1)-- (2,0);
\draw [thick] (2,0)-- (2,-1);
\draw [thick] (2,-1)-- (0.4,-0.65);
\draw [thick] (0.4,-0.65)-- (1,1);
\draw [thick] (4,-1)-- (4,0);
\draw [thick] (4,0)-- (5,1);
\draw [thick] (5,1)-- (5.72,-0.63);
\draw [thick] (5.72,-0.63)-- (4,-1);
\draw [thick] (2,3)-- (2,4)node[above]{4};
\draw [thick] (4,3)-- (4,4)node[above]{5};
\draw [thick] (5.72,-0.63)-- (6.38,-1.27)node[right]{7};
\draw [thick] (3,-2)-- (3,-3)node[below]{1};
\draw [thick] (0.4,-0.65)-- (-0.5,-1.09)node[left]{2};
\begin{scriptsize}
\draw [fill=black] (2,3) circle (3pt);
\draw [fill=white] (3,2) circle (2.5pt);
\draw [fill=black] (4,3) circle (3pt);
\draw [fill=white] (3,1) circle (2.5pt);
\draw [fill=black] (2,0) circle (2.5pt);
\draw [fill=white] (1,1) circle (2.5pt);
\draw [fill=white] (1,2) circle (2.5pt);
\draw [fill=white] (5,2) circle (2.5pt);
\draw [fill=white] (5,1) circle (2.5pt);
\draw [fill=black] (4,0) circle (3pt);
\draw [fill=white] (4,-1) circle (2.5pt);
\draw [fill=white] (3,-2) circle (2.5pt);
\draw [fill=white] (2,-1) circle (2.5pt);
\draw [fill=black] (0.4,-0.65) circle (3pt);
\draw [fill=black] (5.72,-0.63) circle (3pt);
\end{scriptsize}
\end{tikzpicture}
    \caption{The on-shell diagrams corresponding to the solutions~(\ref{eq:ls1-sol}) and~(\ref{eq:ls2-sol}) respectively.}
    \label{fig:plabicgraphs}
\end{figure}

%% file: graph3.tex
\begin{tikzpicture}[line cap=round,line join=round,>=triangle 45,x=1cm,y=1cm,scale=0.5]
\draw [thick] (0.72,2.46)-- (0.72,1.4);
\draw [thick] (0.72,1.4)-- (1.637986928011505,0.87);
\draw [thick] (1.637986928011505,0.87)-- (2.55597385602301,1.4);
\draw [thick] (2.55597385602301,1.4)-- (2.5559738560230105,2.46);
\draw [thick] (2.5559738560230105,2.46)-- (1.6379869280115058,2.99);
\draw [thick] (1.6379869280115058,2.99)-- (0.72,2.46);
\draw [thick] (1.6379869280115058,2.99)-- (2.5559738560230105,2.46);
\draw [thick] (2.5559738560230105,2.46)-- (3.3437073710290477,3.169278442740389);
\draw [thick] (3.3437073710290477,3.169278442740389)-- (2.9125665293687,4.1376366278415455);
\draw [thick] (2.9125665293687,4.1376366278415455)-- (1.858373320278331,4.026836456777834);
\draw [thick] (1.858373320278331,4.026836456777834)-- (1.6379869280115058,2.99);
\draw [thick] (3.3437073710290477,3.169278442740389)-- (2.5559738560230105,2.46);
\draw [thick] (2.5559738560230105,2.46)-- (2.55597385602301,1.4);
\draw [thick] (2.55597385602301,1.4)-- (3.6,0.92);
\draw [thick] (3.6,0.92)-- (4.44,2.76);
\draw [thick] (4.44,2.76)-- (3.3437073710290477,3.169278442740389);
\draw [thick] (0.72,2.46)-- (-0.02,2.96)node[left]{\small{3}};
\draw [thick] (0.72,1.4)-- (-0.04,1)node[left]{\small{2}};
\draw [thick] (1.637986928011505,0.87)-- (1.46,-0.1)node[below]{\small{1}};
\draw [thick] (3.6,0.92)-- (3.88,-0.08)node[below]{\small{7}};
\draw [thick] (4.44,2.76)-- (5.34,3.28)node[right]{\small{6}};
\draw [thick] (2.9125665293687,4.1376366278415455)-- (3.24,5.28)node[above]{\small{5}};
\draw [thick] (1.858373320278331,4.026836456777834)-- (1.32,5.18)node[above]{\small{4}};
\node at (1.6,1.9) {\tiny{EF}};
\node at (3.4,2.1) {\tiny{AB}};
\node at (2.5,3.4){\tiny{CD}};
\end{tikzpicture}

%% file: graph4.tex
\begin{tikzpicture}[line cap=round,line join=round,>=triangle 45,x=1cm,y=1cm, scale=0.5]
\draw [thick](-4,3)-- (-5,2);
\draw [thick](-5,2)-- (-4.357960478079794,0.7399264893298994);
\draw [thick](-4.357960478079794,0.7399264893298994)-- (-2.961158231412374,0.9611582314123733);
\draw [thick](-2.961158231412374,0.9611582314123733)-- (-2.739926489329899,2.357960478079794);
\draw [thick](-2.739926489329899,2.357960478079794)-- (-4,3);
\draw [thick](-2.961158231412374,0.9611582314123733)-- (-1.6881939697252004,0.3162009716857487);
\draw [thick](-1.6881939697252004,0.3162009716857487)-- (-0.6814355750370342,1.32755917387316);
\draw [thick](-0.6814355750370342,1.32755917387316)-- (-1.3321889303476389,2.597570177352593);
\draw [thick](-1.3321889303476389,2.597570177352593)-- (-2.741135016910796,2.3711219414018325);
\draw [thick](-2.741135016910796,2.3711219414018325)-- (-2.961158231412374,0.9611582314123733);
\draw [thick](-2.739926489329899,2.357960478079794)-- (-1.3321889303476389,2.597570177352593);
\draw [thick](-1.3321889303476389,2.597570177352593)-- (-1.5717986296204378,4.005307736334853);
\draw [thick](-1.5717986296204378,4.005307736334853)-- (-2.979536188602698,3.765698037062054);
\draw [thick](-2.979536188602698,3.765698037062054)-- (-2.739926489329899,2.357960478079794);
\draw [thick](-2.979536188602698,3.765698037062054)-- (-3.5546463900092493,4.240984098451268)node[above]{\small{4}};
\draw [thick](-1.5717986296204378,4.005307736334853)-- (-1.2172199945133375,4.589853709719314)node[right]{\small{5}};
\draw [thick](-0.6814355750370342,1.32755917387316)-- (0,1)node[right]{\small{6}};
\draw [thick](-1.6881939697252004,0.3162009716857487)-- (-1.5835330863447863,-0.36409477028694137)node[below]{\small{7}};
\draw [thick](-4.357960478079794,0.7399264893298994)-- (-4.897794393391228,0.1417661660517256)node[below]{\small{1}};
\draw [thick](-5,2)-- (-5.66530753818093,2.0256620668991747)node[left]{\small{2}};
\draw [thick](-4,3)-- (-4.0081768846577095,3.6479057592955897)node[above]{\small{3}};
\node at (-2.1,3.2) {\tiny{EF}};
\node at (-1.9,1.5) {\tiny{CD}};
\node at (-3.8,1.8){\tiny{AB}};
\end{tikzpicture}

%% file: graph5.tex
\scalebox{0.7}{
\begin{tikzpicture}[line cap=round,line join=round,>=triangle 45,x=1cm,y=1cm, scale=0.6]
\draw [thick](0.02,3.07)-- (-1.74,3.03);
\draw [thick](-1.74,3.03)-- (-2.245827649448101,1.3437798515455324);
\draw [thick](-2.245827649448101,1.3437798515455324)-- (-0.7984463292564951,0.34163848728577717);
\draw [thick](-0.7984463292564951,0.34163848728577717)-- (0.6019121707517134,1.4085012110955273);
\draw [thick](0.6019121707517134,1.4085012110955273)-- (0.02,3.07);
\draw [thick](-2.245827649448101,1.3437798515455324)-- (-3.66,0.25);
\draw [thick](-3.66,0.25)-- (-3.0567568340909785,-1.4329543913892941);
\draw [thick](-3.0567568340909785,-1.4329543913892941)-- (-1.269759703526212,-1.3792975552382396);
\draw [thick](-1.269759703526212,-1.3792975552382396)-- (-0.7685779049476742,0.33681858462119063);
\draw [thick](-0.7685779049476742,0.33681858462119063)-- (-2.245827649448101,1.3437798515455324);
\draw [thick](0.6019121707517134,1.4085012110955273)-- (-0.7984463292564951,0.34163848728577717);
\draw [thick](-0.7984463292564951,0.34163848728577717)-- (-1.269759703526212,-1.3792975552382396);
\draw [thick](-1.269759703526212,-1.3792975552382396)-- (0.66,-2.75);
\draw [thick](0.66,-2.75)-- (2.7,-0.09);
\draw [thick](2.7,-0.09)-- (0.6019121707517134,1.4085012110955273);
\draw [thick](-1.74,3.03)-- (-2.2,3.91)node[above]{5};
\draw [thick](0.02,3.07)-- (0.52,4.01)node[above]{6};
\draw [thick](2.7,-0.09)-- (3.48,0.37)node[right]{7};
\draw [thick](2.7,-0.09)-- (3.44,-0.57)node[right]{1};
\draw [thick](0.66,-2.75)-- (0.64,-3.69)node[below]{2};
\draw [thick](-3.0567568340909785,-1.4329543913892941)-- (-3.96,-2.11)node[left]{3};
\draw [thick](-3.66,0.25)-- (-4.58,0.43)node[left]{4};
\node at (-0.8,2) {EF};
\node at (-2.3,-0.2) {CD};
\node at (0.5,-0.4){AB};
\end{tikzpicture}
}

%% file: graph6.tex
\begin{tikzpicture}[line cap=round,line join=round,>=triangle 45,x=1cm,y=1cm,scale=1]
\draw [thick](-2.9,0.5)-- (-2.765187568485854,-0.154144767737979);
\draw [thick](-2.765187568485854,-0.154144767737979)-- (-2.1312753661150197,-0.36446616116976854);
\draw [thick](-2.1312753661150197,-0.36446616116976854)-- (-1.6321755952583312,0.07935721313642133);
\draw [thick](-1.6321755952583312,0.07935721313642133)-- (-1.7669880267724771,0.7335019808744004);
\draw [thick](-1.7669880267724771,0.7335019808744004)-- (-2.400900229143311,0.9438233743061899);
\draw [thick](-2.400900229143311,0.9438233743061899)-- (-2.9,0.5);
\draw [thick](-2.400900229143311,0.9438233743061899)-- (-2.9,0.5);
\draw [thick](-2.9,0.5)-- (-3.6613815892459978,0.6468782065697603);
\draw [thick](-3.6613815892459978,0.6468782065697603)-- (-3.6613815892459978,1.3289373728119938);
\draw [thick](-3.6613815892459978,1.3289373728119938)-- (-2.812773091712056,1.6223814327069082);
\draw [thick](-2.812773091712056,1.6223814327069082)-- (-2.400900229143311,0.9438233743061899);
\draw [thick](-3.6613815892459978,0.6468782065697603)-- (-3.605865145482095,-0.02725003913477277);
\draw [thick](-3.605865145482095,-0.02725003913477277)-- (-2.765187568485854,-0.154144767737979);
\draw [thick](-2.765187568485854,-0.154144767737979)-- (-2.9,0.5);
\draw [thick](-2.9,0.5)-- (-3.6613815892459978,0.6468782065697603);
\draw [thick](-3.6613815892459978,1.3289373728119938)-- (-4.020590426066303,1.5343373643633444)node[left]{\small{6}};
\draw [thick](-3.6613815892459978,0.6468782065697603)-- (-4.10586966437627,0.6272763750664248)node[left]{\small{5}};
\draw [thick](-3.605865145482095,-0.02725003913477277)-- (-3.8655372655027262,-0.26427929817413726)node[left]{\small{4}};
\draw [thick](-2.1312753661150197,-0.36446616116976854)-- (-1.9661360485989174,-0.7061808057803289)node[below]{\small{3}};
\draw [thick](-1.6321755952583312,0.07935721313642133)-- (-1.1676122716965,0.20863284154476955)node[right]{\small{2}};
\draw [thick](-1.7669880267724771,0.7335019808744004)-- (-1.4,1)node[right]{\small{1}};
\draw [thick](-2.812773091712056,1.6223814327069082)-- (-2.632864639022295,1.976238871969536)node[above]{\small{7}};
\node at (-3.3,0.2) {\tiny{EF}};
\node at (-3.1,1.1) {\tiny{CD}};
\node at (-2.3,0.4){\tiny{AB}};
\end{tikzpicture}

%% file: graph7.tex
\scalebox{0.8}{
\begin{tikzpicture}[line cap=round,line join=round,>=triangle 45,x=1cm,y=1cm, scale=1.2]
\draw [thick](-2,3)-- (-1,3);
\draw [thick](-1,3)-- (-1,4);
\draw [thick](-1,4)-- (-2,4);
\draw [thick](-2,4)-- (-2,3);
\draw [thick](-1,3)-- (0,3);
\draw [thick](0,3)-- (0,4);
\draw [thick](0,4)-- (-1,4);
\draw [thick](-1,4)-- (-1,3);
\draw [thick](0,3)-- (-1,3);
\draw [thick](-1,3)-- (-1,2);
\draw [thick](-1,2)-- (0,2);
\draw [thick](0,2)-- (0,3);
\draw [thick](-2,4)-- (-2.36,4.35)node[left]{\small{2}};
\draw [thick](-1,4)-- (-0.98,4.49)node[above]{\small{3}};
\draw [thick](0,4)-- (0.4,4.37)node[right]{\small{4}};
\draw [thick](0,3)-- (0.5,2.99)node[right]{\small{5}};
\draw [thick](0,2)-- (0.4,1.65)node[below]{\small{6}};
\draw [thick](-1,2)-- (-1.36,1.61)node[below]{\small{7}};
\draw [thick](-2,3)-- (-2.38,2.67)node[left]{\small{1}};
\node at (-1.5,3.5) {\tiny{AB}};
\node at (-0.5,3.5) {\tiny{CD}};
\node at (-0.5,2.5){\tiny{EF}};
\end{tikzpicture}
}

%% file: graph8.tex
\scalebox{0.8}{
\begin{tikzpicture}[line cap=round,line join=round,>=triangle 45,x=1cm,y=1cm, scale=1.2]
\draw [thick](-2,0)-- (-1.5,0.53);
\draw [thick](-1.5,0.53)-- (-1,0);
\draw [thick](-1,0)-- (-1,-1);
\draw [thick](-1,-1)-- (-2,-1);
\draw [thick](-2,-1)-- (-2,0);
\draw [thick](-1,0)-- (0,0);
\draw [thick](0,0)-- (0,-1);
\draw [thick](0,-1)-- (-1,-1);
\draw [thick](-1,-1)-- (-1,0);
\draw [thick](-1,0)-- (-1,1);
\draw [thick](-1,1)-- (0,1);
\draw [thick](0,1)-- (0,0);
\draw [thick](0,0)-- (-1,0);
\draw [thick](-1.5,0.53)-- (-1.5,1)node[above]{\small{3}};
\draw [thick](-2,0)-- (-2.413125097836932,0.3692959877502467)node[left]{\small{2}};
\draw [thick](-2,-1)-- (-2.436917859450033,-1.3041282457045353)node[below]{\small{1}};
\draw [thick](0,-1)-- (0.2675260439057988,-1.2644736430160335)node[below]{\small{7}};
\draw [thick](0,0)-- (0.5,0)node[right]{\small{6}};
\draw [thick](0,1)-- (0.37062801089590386,1.2892827701234917)node[above]{\small{5}};
\draw [thick](-1,1)-- (-1.3424508252473797,1.3606610549627953)node[above]{\small{4}};
\node at (-0.5,-0.5) {\tiny{EF}};
\node at (-0.5,0.5) {\tiny{CD}};
\node at (-1.5,-0.4){\tiny{AB}};
\end{tikzpicture}
}

%% file: graph9.tex
\scalebox{0.6}{
\begin{tikzpicture}[line cap=round,line join=round,>=triangle 45,x=1cm,y=1cm]
\draw [thick](0.1,3)-- (-1.9,3);
\draw [thick](-1.9,3)-- (-2.5180339887498944,1.0978869674096934);
\draw [thick](-2.5180339887498944,1.0978869674096934)-- (-0.9,-0.07768353717525289);
\draw [thick](-0.9,-0.07768353717525289)-- (0.7180339887498948,1.0978869674096927);
\draw [thick](0.7180339887498948,1.0978869674096927)-- (0.1,3);
\draw [thick](-2.5180339887498944,1.0978869674096934)-- (-0.9,-0.07768353717525289);
\draw [thick](-0.9,-0.07768353717525289)-- (-2.163828828317477,-1.4107036040910799);
\draw [thick](-2.163828828317477,-1.4107036040910799)-- (-3.7511855595870887,-0.17221648408951595);
\draw [thick](-3.7511855595870887,-0.17221648408951595)-- (-2.5180339887498944,1.0978869674096934);
\draw [thick](-0.9,-0.07768353717525289)-- (0.3480323728124578,-1.3758166429642753);
\draw [thick](0.3480323728124578,-1.3758166429642753)-- (1.952832584645472,-0.2419904063431252);
\draw [thick](1.952832584645472,-0.2419904063431252)-- (0.7180339887498948,1.0978869674096927);
\draw [thick](0.7180339887498948,1.0978869674096927)-- (-0.9,-0.07768353717525289);
\draw [thick](-1.9,3)-- (-1.9,4)node[above]{4};
\draw [thick](0.1,3)-- (0.1,4)node[above]{5};
\draw [thick](-2.5180339887498944,1.0978869674096934)-- (-3.5244203122628583,1.7988968195749448)node[left]{3};
\draw [thick](-3.7511855595870887,-0.17221648408951595)-- (-4.762907432264424,-0.17221648408951595)node[left]{2};
\draw [thick](-2.163828828317477,-1.4107036040910799)-- (-2.146385347754075,-2.335208073951402)node[below]{1};
\draw [thick](0.3480323728124578,-1.3758166429642753)-- (0.40036281450266475,-2.335208073951402)node[below]{7};
\draw [thick](1.952832584645472,-0.2419904063431252)-- (3.0517718601398185,-0.2594338869065275)node[right]{6};
\node at (0.5,-0.3) {EF};
\node at (-0.8,1.5) {CD};
\node at (-2.4,-0.3){AB};
\end{tikzpicture}
}